# Spherically Symmetric Chromostatic Condensates as an Introduction to the Strong Conjecture for Color Confinement

Dennis Sivers
Portland Physics Institute
Portland, OR 97239

Spin Physics Center
University of Michigan
Ann Arbor, MI 48109

Abstract

Nonlinearities imbedded in the Lagrange density for non-Abelian gauge theories produce solutions to the Yang-Mills Maxwell equations that describe spatially extended chromostatic condensates. For solutions in spherically symmetric SU(2) the topological structures separating such condensates provide a specific solitonic description of color confinement.

## The Strong Conjecture

*The confinement mechanism for QCD involves a domain wall of topological (CP-odd) charge separating the interior volume of hadrons from an exterior volume.*

To explain the consequences of this conjecture we describe spherically symmetric chromostatic condensates that are consistent with the interior volume of a hadron and other condensates that are consistent with an exterior vacuum volume. We then demonstrate how the Yang-Mills Maxwell equations can connect the two volumes with a soliton domain wall. The preliminary phenomenological description described here does not deal explicitly either with charged fermions or with the quantization of non-Abelian dynamics.

Section 1: Highlights in the Study of Confinement.

The study of quark confinement predates the formulation of QCD. The MIT bag model [1,2] incorporates the confinement of quarks by postulating a boundary condition and a bag constant. Several variations of the original bag model exist and these models have all been able to provide useful non-gluonic descriptions for important aspects of hadronic structure.[3] The SLAC Bag model [4] attempts to reconcile field theory with the bag model concept and exposes many of the challenges involved in the study of color confinement. A useful starting point in this study involves the topological properties shown in Fig. 1 that are the consequences of color confinement. The topological structure of this picture already provides important information about these consequences. The interior volume in this sketch has the topology of a 3-torus, T3. This volume contains the color sources and the fields carrying color charges. The exterior volume in the figure is also a 3-torus, this time a "vacuum" state from which color is excluded. The intermediate, transition, volume shown has the topology of a 2-sphere, S2, extended into a shell, S2xI. The contrasts with the MIT bag model, in which the boundary displays the simpler topology of a 2-sphere, S2. By extending the bag surface into a domain wall we give room for the gradients of the Yang-Mills Maxwell equations to activate the non-perturbative dynamics required for confinement. The identification of the three spherically-symmetric volumes shown in the picture is, however, consistent with the other global features of bag models. Color confinement requires that interior volume exists in an overall color-singlet state and that color charges in the interior volume find it impossible to escape. However, the intermediate volume, as shown in the figure, displays an extended volume that is more compatible with the quantum nature of the problem and, most importantly, is compatible with the hypothesis that the confinement mechanism is an intrinsic feature of QCD and not an added gimmick. In addition, the exterior volume displayed here is also not entirely the same as the exterior of a bag model. Because we are dealing with a gauge theory, the gauge connection has the power to impose a chiral structure from the interior volume directly connecting to a chiral structure on the exterior volume. The importance of the topological labels attached to these volumes involves the boundary conditions involved with this chiral structure as will explained later in the context of chromostatic condensates and domain walls.

The importance of finding a mechanism or an emergent structure in a quantum field theory that explains the confinement of color can be found by studying the lucid arguments in any number of excellent textbooks [5]. A truly fundamental understanding of confinement can ultimately provide an essential link between the microscopic perturbative quark-gluon degrees of freedom and the emergent nonperturbative structure that leads to the physically observed strong-interaction sepectrum of color-neutral mesons and baryons and to complex nuclei. The most predictive approach to understanding confinement is based on the regularization of QCD on a Euclidean lattice. The formulation of lattice techniques based on Monte-Carlo integration has tamed the issue of the renormalization of quantum field theories in the infra-red (IR) and strong-coupling (SC) limits. Lattice studies also present a compelling argument that the confinement of color charge in the fundamental representation can be understood in terms of the linearly-rising potential that is found in the so-called "quenched approximation" that neglects the fermion determinant. Detailed calculations of the hadronic spectrum [6] have combined with lattice calculations that illustrate the dressing of light quarks to generate an effective quark mass parameter that approximates the non-relativistic constituent quark mass in the infra-red limit. These calculations have been replicated by momentum space diagram summations and the Schwinger-Dyson equations [7]. Confirmation that lattice calculations represent fundamental *ab initio* calculations of QCD dynamics,[8] can be found in the fact that explicit approximation schemes for the calculation of hadronic distribution functions have now been formulated.

We need to respect these achievements, thus the simple approach presented here does not challenge the power of lattice QCD. Instead, it provides simple illustrations based strictly on the chiral structure imposed by the Yang-Mills Maxwell equations of spherically symmetric SU(2). The confinement of color is a unique property in field theory. Just as taking off a right-handed glove and turning it inside out creates a left-handed glove, the separation between an interior volume and an exterior volume of a one-particle state can display a distinctive chiral signature. The importance of this signature is under-scored by models of color confinement based on the dual-Meissner effect [9,10]. As described, in detail, by Ripka [11] these models use the magnetic dual of the Landau-Ginzberg Lagrange density to postulate a magnetically charged condensate that forms the exterior, "vacuum", which surrounds a hadronic one-particle system and provides the color-confinement mechanism. In this article we will demonstrate that a sterile vacuum state with a vanishing field strength tensor and no electric color currents can also have a chiral structure imposed by the gauge connection that also leads to color confinement. Based directly on the Yang-Mills Maxwell equations, we will use SU(2) chromostatic condensates to illustrate the **Strong Conjecture** cited in the abstract,

*The confinement mechanism for QCD involves a domain wall of topological (CP- odd) charge separating the interior volume of hadrons from an exterior volume.*

The strong conjecture postulates a topological soliton to impose the separation required for the color confinement of quarks and gluons. The arguments underlying this conjecture will be explained here in terms of the domain structure found in the Ralston-Sivers [12-14] "field-strength-description" of spherically symmetric SU(2). The connection of these arguments to SU(N) and the inclusion of fermions in the fundamental representation of the gauge group will be deferred to a later discussion. The basic argument presented here is that the classical SU(2) Yang-Mills Maxwell equations are relevant for confinement and can be explained by solving these equations with confining boundary conditions applied to the volumes shown in Fig. 1. Section 2 in this paper discusses the assumptions underlying the formulation of the strong conjecture. Specifically, it considers the conditions required for the existence of topological solitons in field theories and discusses the emergence of longitudinal and transverse chiral structures in three spatial dimensions for static solutions to the field equations with electric color sources. Section 3 then presents the static condensates found in spherically-symmetric SU(2) chromostatics and classifies their properties in terms of the color-averaged Lorentz invariant field strength densities,

$$L(0^{++}) = E_i^a E_i^a - B_i^a B_i^a \quad (J^{PC} = 0^{++})$$
$$L(0^{-+}) = E_i^a B_i^a \quad (J^{PC} = 0^{-+}),$$
(1.1)

the static color-charge currents, and importantly, the chiral properties created by the gauge-covariant derivative. Section 4. introduces solutions to the Yang-Mills Maxwell equations that create the domain walls of topological charge that are the generalization of kink-solutions in 1+1 dimensional theories. These are regions where the pseudoscalar Lorentz- invariant $L(0^{-+}) \neq 0$, that can separate the internal and external chromostatic condensates into color-confining emergent structures. Section 5. provides some conclusions and discusses the status of the strong conjecture prize offered to anyone who can disprove the strong conjecture.

Section 2. Introducing the Postulates of the Strong Conjecture

A proposal that the confinement mechanism in a non-Abelian gauge theory involves an extended solitonic structure such as a domain wall includes consideration of the possibilities for the existence of such solutions to the Yang-Mills Maxwell equations. Derrick [15] addressed this problem by studying the scale-invariance of stable solutions to the field equations in different dimensions, $n_D$, of Euclidean space-time. Consider an SU(N) gauge theory coupled to electrically charged scalar fields with a Euclidean Lagrangian density given by,

$$L^{n_D} = -\frac{1}{2} G_{\mu\nu}^a G_a^{\mu\nu} + \frac{1}{2} (^A D_\mu \phi)_a^\dagger (^A D^\mu \phi)^a - V(\phi).$$
(2.1)

In this equation $^A D_\mu$ denotes the gauge-covariant derivative specified by the gauge connection $A_\mu^a$ and $\phi_a$ is a scalar field carrying a color charge. For convenience, Derrick removed the static electric fields, $E_i^a = G_{oi}^a$ from consideration by setting $A_0^a = 0$. A stable solution to these equations is then defined by the Euclidean-space integrals,

$$I_B(A) = \frac{1}{2} \int d^{n_D} x (B_i^a B_i^a)$$

$$I_K(A, \phi) = \frac{1}{2} \int d^{n_D} x (^A D_i \phi)_a^\dagger (^A D_i \phi)_a \qquad (2.2)$$

$$I_V(\phi) = \int d^{n_D} x V(\phi).$$

Because these integrals are defined to be non-negative, we then specify the scaling properties of a static solution for different dimensions, $n_D$, of space-time. By inserting a label on the gauge connection $\bar{A}_i^a(x)$ and the scalar field $\bar{\phi}^a(x)$ it is possible to define scaled solution by simple dimensional labels

$$\alpha_{i,\lambda}^a(x) = \lambda \bar{A}_i^a(\lambda x)$$
$$f_\lambda^a(x) = \bar{\phi}_a(\lambda x). \qquad (2.3)$$

This scaling has an impact on the static energies,

$$H_\lambda^S = I_B(\alpha_\lambda) + I_K(\alpha_\lambda, f_\lambda) + I_V(f_\lambda)$$
$$H_\lambda^S = \lambda^{4-n_D} I_B(\bar{A}) + \lambda^{2-n_D} I_K(\bar{A}, \bar{\phi}) + \lambda^{-n_D} I_V(\bar{\phi}). \qquad (2.4)$$

The sum $H_\lambda^S$ will be stationary at $\lambda = 1$ if and only if

$$0 = (n_D - 4) I_B(\bar{A}) + (n_D - 2) I_K(\bar{A}, \bar{\phi}) + n_D I_V(\bar{\phi}) \qquad (2.5)$$

For the SU(N) gauge theory with scalar fields in $n_D = 2$ Euclidean dimensions this leads to the constraint

$$-2 I_B(\bar{A}) + 2 I_V(\bar{\phi}) = 0 \qquad (2.6)$$

This potential stable solution in 2 dimensions therefore requires a balance between the magnetic and potential energies. For a soliton in $n_D = 3$ Euclidean dimensions the scaling conditions give

$$-I_B(\bar{A}) + I_K(\bar{A}, \bar{\phi}) + 3 I_V(\bar{\phi}) = 0 \qquad (2.7)$$

Thus, a potential soliton has two positive semidefinite terms to be balanced against the negative semidefinite term involving the magnetic field strength density. For a stable solution in $n_D = 4$ Euclidean dimensions we have the condition

$$2 I_K(\bar{A}, \bar{\phi}) + 4 I_V(\bar{\phi}) = 0. \qquad (2.8)$$

This final condition contains only positive coefficients and thus requires that each nonnegative integral vanish so that there can be no contribution from charged scalar fields unless their charge vanishes and there are no gradients. It therefore allows only a cosmological constant not a local field. Separately, it requires that $I_B(\bar{A})$ be independently scale invariant.

The dimensional analysis performed by Derrick can also be used in the study of confined systems in 3+1 dimensions with a Minkowski metric. For this purpose we will follow the conditions described above and not consider any scalar fields except to impose or remove a cosmological term [16] and will instead start the analysis with an integral of the form,

$$I_G(A_\mu^a, J_\mu^a) = \oint_{conf} dt d^3\vec{x} F(B_i^a B_i^a - E_i^a E_i^a, E_i^a B_i^a; ...) \tag{2.9}$$

involving the Lorentz-invariant color averaged densities. The integrand here defines a measure of pressure for a confined system of color sources that contains Lorentz-invariant field-strength densities and electric current sources that are also confined. Initially, we will restrict attention to the two densities shown. Within this framework, we will look for chiral structures that allow for SU(N) gluons to be confined while remaining massless. Two, specialized versions of this integral, are particularly useful because they connect to the solitons found in 2 or 3 dimensions for theories with color-electric charged scalars. The first case is important for the study of confinement in the rest-frame of a confined system. It assumes the field-strength densities are spherically symmetric so that we can write:

$$I_G^S(A_\mu^a; J_\mu^a) = 2\pi \int dt \int_{conf} r^2 dr F^S(B_i^a B_i^a(r,t) - E_i^a E_i^a(r,t); E_i^a B_i^a(r,t)) \tag{2.10}$$

where it is assumed we have a solution to the Yang-Mills Maxwell equations with spherical symmetry for color electric sources and confining boundary conditions. The condition that we can construct a stable confined system that would produce a pole in the hadronic S-matrix is crucial to the strong conjecture described here. We will show that this global system should have CP-odd solitons that correspond to those found in 2 dimensional Euclidean space with scalar fields. The second case uses cylindrical symmetry for the field-strength densities,

$$I_G^C(A_\mu^a, J_\mu^a) = \pi \int dt \iint_{conf} \rho d\rho dz F^C(B_i^a B_i^a - E_i^a B_i^a; E_i^a B_i^a)(\rho, z, t) \tag{2.8}$$

In this integral we have condensed the form to not show the $\rho, z, t$ dependence of the individual terms. This integral is appropriate for the central region of a rapidly expanding jet or for a massive, large-orbital-angular-momentum state located on a linearly-rising Regge trajectory where the Yang-Mills Maxwell equations are formulated in the co-rotating frame. The solitons in this system should correspond to those found in 3-dimensional Euclidean space with scalar fields. Note that these arguments do not depend on the number of colors in SU(N).

In either case (2.7) or (2.8) above, the study of color confinement involves the chiral structure of extended systems. For clarity and simplicity, we will only define chiral systems in terms of 3-dimensional properties and restrict attention to the gauge group SU(2). Let $v_i$ denote a vector in 3-space and $\sigma_i$ denote an axial vector so that the application of the parity operator can be given by:

$$P(v_i, \sigma_i)P^{-1} = (-v_i, \sigma_i) \tag{2.11}$$

In addition, we can define the Hodge-dual of the parity operator, $^*P$, in terms of the transformation given by

$$^*P(v_i, \sigma_i)(^*P)^{-1} = (v_i, -\sigma_i) \tag{2.12}$$

Adjoint charges in SU(2) map onto the SO(3) of rotations so it is possible to define vectors and axial vectors in the adjoint representation of SU(2) color space in a similar manner so it is possible to define color vectors, $v_a$, and axial vectors $\tau_a$, in a similar manner to the classification used for 3-space,

$$C(v_a, \tau_a)C^{-1} = (-v_a, \tau_a), \tag{2.13}$$

For SU(N) charges in the adjoint representation we can also use the root vectors to define a charge conjugation operator with these properties but we will not explore this freedom at this time. We will, however, use the fact that Hodge duality in an adjoint representation of color space can be treated in a similar fashion with Hodge duality in 3-space. For a full discussion of charge conjugation we would have to introduce SU(2) charges in the fundamental representation. In order to use the Yang-Mills Maxwell equations for the study of extended structure of adjoint color charges, it is then convenient to define longitudinal and transverse components for the basis vectors in the direct product space in a congruent manner such that,

$$\begin{aligned} P(L_{ia}, \delta_{ia}^T, \varepsilon_{ia}^T)P^{-1} &= \pm(-L_{ia}, -\delta_{ia}^T, \varepsilon_{ia}^T) \\ C(L_{ia}, \delta_{ia}^T, \varepsilon_{ia}^T)C^{-1} &= \pm(-L_{ia}, -\delta_{ia}^T, \varepsilon_{ia}^T). \end{aligned} \tag{2.14}$$

These equations demonstrate that chirality can be related both to parity and to adjoint charge conjugation. The overall $\pm$ sign indicates that in extended systems these global operations can be combined with rotations and gauge transformations. The equations in (2.12) lead to

$$PC(L_{ia}, \delta_{ia}^T, \varepsilon_{ia}^T)C^{-1}P^{-1} = \pm(L_{ia}, \delta_{ia}^T, \varepsilon_{ia}^T) \tag{2.15}$$

and, in discussing extended structures, it is clear that CP-odd does not imply CP-violating but can, instead, imply CP separation. In discussing the field-strength densities $E_i^a(x), B_i^a(x)$ we use

$$P(E_i^a(\vec{x},t), B_i^a(\vec{x},t))P^{-1} = (E_i^a(-\vec{x},t), B_i^a(-\vec{x},t)) = (-E_i^a(\vec{x},t), B_i^a(\vec{x},t)) \quad (2.16)$$

Given the fact that the Lagrangian density for the gauge sector is self-dual and frequently written in the form,

$$L_G(x) = G \wedge {}^*G \quad (2.17)$$

a good deal of attention is given to the self-dual sector

$$E_i^a(x) = \pm B_i^a(x) \quad (2.18)$$

of the non-Abelian field equations. However, (2.18) is not consistent with real field-strengths in Minkowski space and diverts attention from the chiral properties in 3-space that are crucial to the differential geometry involved in the confinement mechanism of color-electric charge. The exact definition of a gauge connection, $A_\mu^a(\vec{x},t)$ with sufficient topological structure and the existence of a set of non-local color charges defined by a covariantly conserved current $J_\mu^a(\vec{x},t)$ are required to determine the domain structure of a confined system. The field-strength description of spherically-symmetric SU(2) developed by Ralston and Sivers has the necessary tools to define the problem appropriately.

Section 3. Spherical Condensates in SU(2) Chromostatics.

The Ralston-Sivers field-strength description for spherically symmetric SU(2) is presented in more detail in Ref. [17]. In this section, the goal is to use this formalism to present the chiral structure of extended condensates in spherical SU(2). The electric and magnetic field-strength densities generated by the gauge connection, $A_\mu^a(r,t)$,

$$A_0^a(r,t) = A_0(r,t)\hat{r}_a$$
$$A_i^a(r,t) = A_1(r,t) + \frac{a(r,t)}{r}\varepsilon_{ia}^A(\omega(r,t)) - \frac{1}{r}\varepsilon_{ia}^A(0) \quad (3.1)$$

are given by,

$$E_L = \partial A_1/\partial t - \partial A_0/\partial r \qquad B_L = \frac{a^2-1}{r^2}$$

$$E_S = \frac{a}{r}(A_0 - \partial\omega/\partial t) \qquad B_S = -\frac{1}{r}\partial a/\partial r \quad (3.2)$$

$$E_A = \frac{1}{r}\partial a/\partial t \qquad B_A = -\frac{a}{r}(A_1 - \partial\omega/\partial r).$$

In these equations, the basis vectors for spherical tensors are,

$$\rho_{ia} = \hat{r}_i \hat{r}_a$$
$$\delta_{ia}^T = (\hat{\theta}_i \hat{\theta}_a + \hat{\phi}_i \hat{\phi}_a) \tag{3.3}$$
$$\varepsilon_{ia}^T = (\hat{\phi}_i \hat{\theta}_a - \hat{\theta}_i \hat{\phi}_a)$$

Note that with this convention the longitudinal tensor has trace 1 while the transverse tensors have trace 2. The definition of these tensors combining spatial and SU(2) color coordinates is explained if Fig. 2 and repeated in the appendix. In all these equations, the gauge coupling has been set to unity and we have expanded the field-strength densities in terms of gauge-covariant tensors given by

$$^A D_i^{ab} \hat{r}_b = \frac{a}{r} \varepsilon_{ia}^S(\omega) = \frac{a}{r}[\delta_{ia}^T \cos(\omega) - \varepsilon_{ia}^T \sin(\omega)]. \tag{3.4}$$

$$-i[\hat{r},{}^A D_i \hat{r}]^a = \frac{a}{r} \varepsilon_{ia}^A(\omega) = \frac{a}{r}[\delta_{ia}^T \sin(\omega) + \varepsilon_{ia}^T \cos(\omega)]. \tag{3.5}$$

With the gauge-covariant SU(2) current, $j_\mu^a(r,t)$, expanded in terms of the same tensor structure

$$j_0^a(r,t) = \frac{1}{r^2} J_0(r,t) \hat{r}_a$$
$$j_i^a(r,t) = \frac{1}{r^2} J_1(r,t)\rho_{ia} + j_S(r,t)\varepsilon_{ia}^S(\omega(r,t)) + j_A(r,t)\varepsilon_{ia}^A(\omega(r,t)) \tag{3.6}$$

the generalization of Maxwell's equations to the SU(2) system $^A D_{ab}{}^\mu G_{\mu\nu}^b = j_\nu^a$ then gives the set of equations,

$$\frac{\partial}{\partial r}(r^2 E_L) = J_0 - 2arE_S$$
$$\frac{\partial}{\partial t}(r^2 E_L) = J_1 - 2arB_A, \tag{3.7}$$

for the longitudinal components of the classical SU(2) current and,

$$\frac{\partial}{\partial t}(arE_S) - \frac{\partial}{\partial r}(arB_A) = arj_S \tag{3.8}$$

$$a(-\frac{\partial^2}{\partial t^2} + \frac{\partial^2}{\partial r^2})a - a^2 \frac{(a^2-1)}{r^2} = arj_A + (E_S^2 - B_A^2), \tag{3.9}$$

for the transverse structure. We note here that all the nonlinearities of the non-Abelian Maxwell equations occur in (3.9). The generalization of the Bianchi constraints $^A D_{ab}^\mu {}^*G_{\mu\nu}^b = 0$ provides the additional equations,

$$\frac{\partial}{\partial r}(arE_A) - \frac{\partial}{\partial t}(arB_S) = 0$$
$$-E_L - \frac{\partial}{\partial r}(arE_s) + \frac{\partial}{\partial t}(arB_A) = r^2 E_i^a B_i^a. \tag{3.10}$$

The SU(2) Yang-Mills Maxwell equations provide for an extremely robust set of phenomena in which the function $a(r,t)$ plays many crucial roles. In the limit of spherical symmetry, all of the nonlinear behavior of the non-Abelian dynamics is associated with this function. It sets the normalization for the longitudinal magnetic field strength and for all of the transverse electric and magnetic field-strengths. Because the gauge connection is real analytic the function $a(r,t)$ is also real analytic. This implies that $a(r,t)$ cannot go from $a(r,t)=1$ to $a(r,t)=-1$ without passing through $a(r,t)=0$. Only the longitudinal electric field strength is independent of $a(r,t)$ in spherically symmetric SU(2). The CP-odd density on the RHS of (3.10) is related to the longitudinal divergence of a 2-dimensional axial current,

$$\partial_l K_l(r,t) = r^2 E_i^a B_i^a \qquad (3.11)$$

labeled the topological current, and given by the components,

$$K_0(r,t) = (a^2 - 1)A_1 - a^2 \partial\omega/\partial r$$
$$K_1(r,t) = -(a^2 - 1)A_0 + a^2 \partial\omega/\partial t. \qquad (3.12)$$

Note that this current is independent of the sign of a and is conserved trivially whenever $a(r,t) = \pm 1$. This current thus has the property of being a 2-dimensional dual of the longitudinal gauge connections so that the combinations $A_l + \varepsilon_{lm} K_m$ are gauge-invariant. This allows the gauge-independent transverse components of the field-strength densities,

$$arB_A = -a^2(K_0 + A_1)$$
$$arE_S = a^2(K_1 - A_0). \qquad (3.13)$$

to be closely identified with the chiral structure of the topological domain zones. The topological current defined here is the same current that appears in the Adler-Bell-Jakiw anomaly [18] and, because of this, it must be considered when dealing with charged fermions in the theory.

To focus on the mechanism of confinement, based on the discussion in Section 2, it is useful to simplify further and examine the chromostatic solutions of equations (3.7)-(3.13).

At this point we will begin to write $\partial f(r)/\partial r = f'(r)$ and denote the electric and magnetic field strengths by

$$E_L = -A_0' \qquad E_S = \frac{a}{r}A_0 \qquad E_A = 0$$
$$B_L = \frac{a^2 - 1}{r^2} \qquad B_S = 0 \qquad B_A = \frac{a}{r}(A_1 - \omega'). \qquad (3.14)$$

Then the chromostatic limit of the Yang-Mills Maxwell equations are given by,

$$(r^2 E_L)' = J_0 - 2arE_S$$
$$0 = J_1 - 2arB_A$$
$$(arB_A)' = 2J_1'$$
$$aa'' - a^2 \frac{a^2-1}{r^2} = arj_A + r^2(E_S^2 - B_A^2).$$
(3.15)

These equations verify that the value of this function $a(r)$ can play a major role in classifying the chromostatic condensates. Because the only nonlinearities in these equations involve the function $a(r)$ we will define spherical chromostatic condensates as spherical volumes given by intervals $r \in (R_1, R_2)$ in which $a'(r) = a''(r) = 0$. We will also pay special attention to the values $a = \pm 1$ $a = 0$. Note that the final equation in (3.15) admits solutions that pass from $a(r) = 1$ to $a(r) = -1$ and to $a(r) = 0$. Applying this definition, it is possible to distinguish at least seven distinct classifications of spherically symmetric chromostatic condensates:

1. Sterile chiral vacuum condensates with $a(r) = \pm 1$.

The chiral structure of the chiral vacuum condensates can explain the complications in defining vacuum states in a non-Abelian gauge theory. These "chiral-vacuum" condensates are defined by vanishing of the field-strengths and currents,

$$E_L = E_S = E_A = 0$$
$$B_L = B_S = B_A = 0$$
$$J_0 = J_1 = j_S = j_A = 0.$$
(3.16)

These expressions lead directly to the Lorenz densities

$$L(0^{++}) = 0 \quad L(0^{-+}) = 0$$
(3.17)

However, several examples of a nontrivial gauge connection that leads to these vanishing field-strength densities can still exist. This can be important when the volume is restricted to some $R_0 \geq 0$. Using the source-free version of (3.5) we see that the gauge connection

$$A_i^a(r) = \omega'(r)\rho_{ia} + \frac{a(r)}{r}\varepsilon_{ia}^T(\omega(r)) - \frac{1}{r}\varepsilon_{ia}^A(0)$$
(3.18)

leads to vanishing field-strength densities. Therefore, setting $A_1(r) = \omega'(r)$ leads to a gauge-equivalent infinite class of gauge connections leading to degenerate states dependent on the boundary conditions for $\omega(r)$. The topological current density for these chromostatic condensates is given by

$$K_0(r) = -a^2(r)\omega'(r) \tag{3.19}$$

so topology and chirality both play an important role in their definition. The chirality of each of these condensates is determined by the sign of the function $a(r)$, $a = \pm 1$. It is important for the field-theoretical interpretation of these vacuum condensates [16] that they do not contribute to the cosmological constant. This is the reason we add the adjective "sterile" to their definition. It is crucial to confinement that even a "sterile" vacuum can have chiral structure. This is because the "gluons" that are excitations of the gauge connection must be massless in order for gauge invariance to be compatible with Lorentz invariance. A massless "gluon" can have only two degrees of freedom instead of the three degrees of freedom required for full rotational invariance. Identifying the exterior volume shown in Fig. 1 as a chiral vacuum condensate can while starting from an interior volume of a color glass condensate can, in the language of reference [17], result in either a type-0 or a type-2 solution to that Yang-Mills Maxwell equations with confining boundary conditions.

1. Chiral electric condensates with $a(r) = \pm 1$.

These condensates will have the field-strength densities given by

$$E_L = -A_0' \quad E_S = \pm \frac{A_0}{r} \quad E_A = B_L = B_S = B_A = 0. \tag{3.20}$$

The Lorentz-invariant densities $L(0^{++}) = E_i^a E_i^a - B_i^a B_i^a$ and $L(0^{-+}) = E_i^a B_i^a$ are then given by

$$L(0^{++}) = A_0'^2 + \frac{2}{r^2} A_0^2 \quad L(0^{-+}) = 0. \tag{3.21}$$

Like sterile vacuum condensates, the chirality of these chromostatic spherical condensates is determined the sign of aI and the topological current density is

$$K_0(r) = -a^2 \omega'(r). \tag{3.22}$$

The SU(2) current density that supports the electric condensates is given by

$$J_0(r) = 2ra(r)E_S(r) \tag{3.23}$$

The longitudinal Electric field, $E_L(r)$, is then constrained to be

$$E_L(r) = \pm E_S(r) \tag{3.24}$$

Note that the chiral electric condensates do not have full Lorentz Invariance because under a Lorentz boost the electric fields mix with magnetic fields. Thus, to be part of a Lorentz-invariant emergent structure, chiral electric condensates must be combined with the next type of condensate.

2. Chiral magnetic condensates with $a(r) = \pm 1$.

The chiral magnetic condensates are characterized by the existence of the field strength densities

$$E_L = E_S = E_A = B_L = B_S = 0 \quad B_A = \pm\frac{1}{r}(A_1 - \omega'). \tag{3.25}$$

The Lorentz-invariant densities are given by:

$$L(0^{++}) = -\frac{2}{r^2}(A_1 - \omega')^2 \quad L(0^{-+}) = 0. \tag{3.26}$$

These chiral magnetic condensates illustrate value of the gauge-dependent coordinate basis in that the magnetic field-strength, $B_A(r)$, combines the longitudinal component, $A_1(r)$, with the winding parameter, $\omega(r)$. The topological charge density is given by

$$K_0(r) = -a^2 \omega'(r) \tag{3.27}$$

This extended system requires the SU(2) currents

$$J_1(r) = 2aB_A(r) \quad j_S(r) = \frac{1}{r}(rB_A(r))' \quad j_A(r) = \frac{a}{r}B_A^2 \tag{3.28}$$

Of course, the chirality of the condensates is determined by the sign of the function, $a(r) = \pm 1$, and these systems are required to have nontrivial, finite, extent and can be combined with chiral electric condensates to form the next class of condensates.

3. Chiral –glass condensates with $a(r) = \pm 1$.

Of all the spherically symmetric chromostatic condensates considered here, the chiral-glass condensates are, of course, the most interesting.

More than just an amalgam of electric and magnetic condensates, these extended emergent structures display the level of complexity required to occupy the interior volume of hadrons. This is because a chiral glass condensate can have full Lorentz invariance, including rotational invariance and a meaningful definition of parity. Therefore, a color-neutral set of fermionic fields can be inserted into a chiral glass condensate if they form an irreducible representation of the rotation group. The quantum numbers of the "hadrons" formed in this will then depend on the quantum numbers of the currents. The terminology for this condensate is borrowed from heavy-ion collision physics but does not imply any saturation effects, only the rotational behavior. The field-strength densities for color-glass spherical condensates are given by,

$$E_L = -A_0' \quad E_S = \pm \frac{1}{r} A_0 \quad B_A = \mp \frac{1}{r}(A_1 - \omega')$$
$$B_L = 0 \quad B_S = 0 \quad E_A = 0$$
(3.29)

The Lorentz densities are given by,

$$L(0^{++}) = A_0'^2 + \frac{2}{r^2}[A_0^2 - (A_1 - \omega')^2] \quad L(0^{-+}) = 0.$$
(3.30)

These condensates are supported by a full set of SU(2) electric-color charged currents

$$J_0(r) = (r^2 E_L)' + 2arE_S \quad J_1(r) = 2arB_A$$
$$j_S(r) = \frac{1}{r}(rB_A)' \quad j_A(r) = -\frac{a}{r}(E_S^2 - B_A^2)$$
(3.31)

with a topological charge density specified by

$$K_0(r) = -a^2 \omega'$$
(3.32)

Because the expression for the covariant derivative contains the function $a(r)$,

$$^A D_i^{ab} \hat{r}_b = a(r) \varepsilon_{ia}^S(\omega)$$
(3.33)

the sign of $a(r)$ determines the chirality of the condensate. Chiral glass condensates therefore provide solid scaffolding for the description of the interior volume of hadrons. As we shall discuss further, it is important that this scaffolding carries a chiral signature that distinguishes between electric and magnetic degrees of freedom. The self-duality of the nonAbelian Lagrange density is explicitly broken by the selection of an of an "electric" gauge connection. By bringing in a full set of currents these condensates allow the nonlinearity of the field equations given in (3.15) to select solitonic solutions to confinement with CP-odd domain walls appearing as a consequence of the Yang-Mills Maxwell equations.

4. Chiral 't Hooft-Polyakov condensates with $a(r)=0$.

Building on the seminal discoveries by 't Hooft [19] and Polyakov[20], these condensates have been extensively studied.[21,22] To focus on color confinement the discussion of these chromostatic structures will be restricted here to spherical volumes with a radial interval $r \in (R_0, \infty)$ for some finite value of $R_0$. This restriction is imposed because the parameterization of the gauge connection contains a potential pole at the origin. In order for this pole to vanish for the case $B_L = -\frac{1}{r^2}$ at large values of r, we must have $a(0)=1$. For $B_L = +\frac{1}{r^2}$ at large r, the requirement is $a(0)=-1$. For field strengths in the designated interval the magnitude of the field-strength densities for these condensates are given by,

$$B_L = \pm \frac{1}{r^2} \quad E_L = E_S = E_A = B_S = B_A = 0. \tag{3.34}$$

The Lorentz densities are then given by,

$$L(0^{++}) = -\frac{1}{r^4} \quad L(0^{-+}) = 0. \tag{3.35}$$

These condensates will also be discussed further because of their distinct topological signature required for the vanishing of transverse degrees of freedom with color charge.

$$K_0(r) = A_1 \quad K_1(r) = -A_0 \tag{3.36}$$

The SU(2) "electric" currents for these condensates vanish everywhere

$$J_0(r) = J_1(r) = j_S(r) = j_A(r) = 0 \tag{3.37}$$

Because of the imposition of the condition $a(r)=0$, the chirality of these condensates is necessarily required to be specified by the sign of the longitudinal magnetic field. This condition, $a(r)=0$, completely repels all color charges. In this sense, the 't Hooft-Polyakov condensates provide rotationally-invariant, classical, parameterizations of the dual Meissner effect. This requires a vacuum renormalization of the quantum theory that could involve a fine tuning problem to avoid a very large cosmological constant.

The list of spherically symmetric chromostatic condensates concludes here with two condensates that will not be discussed in further detail at this time because they do not immediately appear to contribute significantly to the concept of color confinement and the strong conjecture. However, their existence signifies an important feature of non-Abelian dynamics that must certainly have profound consequences. There are two questions about the observable universe that are currently unanswered which these exotic condensates preview. Where is the antimatter? Where are the magnetic matter and anti-magnetic matter?

5. Dyonic condensates with $a(r) = 0$.

The field-strength densities for dyonic condenstates are given by,

$$E_L = -A_0' \quad B_L = \pm \frac{1}{r^2} \quad E_S = E_A = B_S = B_A = 0. \tag{3.38}$$

The Lorentz densities are:

$$L(0^{++}) = A_0'^2 - \frac{1}{r^4} \quad L(0^{-+}) = \pm \frac{A_0'}{r^2} \tag{3.39}$$

The chiral signature and SU(2) currents are the same as for the 't-Hooft Polyakov condensates but the magnetic pole at the origin is explicit.

Section 5. Chiral topological condensates with $a(r) = \pm c$, a constant $\neq 0, 1$.

$$E_L = A_0' \quad E_S = \pm \frac{cA_0}{r} \quad B_L = \frac{c^2 - 1}{r^2} \quad B_A = \mp \frac{c(A_1 - \omega')}{r}$$
$$E_A = B_S = 0. \tag{3.40}$$

These condensates may provide clues to discuss strongly interacting matter constrained by extreme conditions involving other forces. For this reason, we will not include Lorentz densities.

All of the above classes of chromostatic condensates are important in their own way. The study of condensates in the static limit includes tools that can convey significant additional dynamical information. For example, in parameterizations of the gauge connection where $A_0(r) = 0$, the time evolution of the composite system can be defined in terms time-directed Wilson lines at $r = 0$, using

$$^A D_o^{ab} = \varepsilon^{abc} A_0(r) \hat{r}_c. \tag{3.41}$$

It s also possible to use Wilson loops to define adiabatic color-flow Poynting vectors

$$P_{ia}^L(r) = E_S(r) B_A(r) \rho_{ia}$$
$$P_{ia}^T(r) = -E_L(r) B_A(r) \varepsilon_{ia}^S(\omega(r)) \tag{3.42}$$

While it is possible these tools to further subdivide the classifications of chromostatic condensates to include additional features of the functional dependence of the gauge-covariant field-strength densities and/or the gauge-covariant color current, at this point it is both appropriate and crucial to turn to the other topological structure specified in the strong conjecture—a domain wall.

Section 4. Domain Walls and Topological Solitons.

The spherically symmetric chromostatic condensates in Section 3 are characterized by distinct values for the Lorentz-invariant densities $L(0^{++})$ and $L(0^{+-})$. Their most important characteristics are determined by the value of the function $a(r)$. For the chiral condensates with $a(r) = \pm 1$, the chiral properties are completely determined by the sign of $a(r)$ as explicitly displayed by the chromostatic Yang-Mills-Maxwell equations given in (3.14). Of these condensates the two sterile vacuum condensates with vanishing currents are candidates the exterior volume shown in Fig. 1. As previously indicated in Section 3, the 't Hoof-Polyakov condensates are also a candidates for the exterior volume of and extended system containing a single hadron.

While the electric and magnetic condensates are interesting in their own right, the only meaningful candidates for the interior volume of an SU(2) hadron are the chiral-glass condensates. In the chiral-glass condensates both the sign of the transverse field strengths and of the classical current densities are required to specify the chirality of the extended system. In the static limit, a single chiral structure extending through the entire interior volume of a hadron is provides for gauge invariance. The observation that massless gluons can only possess two polarization states suggests strongly that the transition volume in Fig. 1 involves a region where the function $a(r)$ changes value. We will describe here two solitonic solutions to the Yang-Mills Maxwell equations that describe that change,

        a) The 't Hooft-Polyakov monopole soliton
        b) The chiral transition soliton.

Both of these solitons involve a domain wall of topological charge. In this context, a spherically-symmetric domain wall can be defined as a spherical volume consisting of an extended region of r in which $a'(r)$ and $a''(r)$ are both nonzero. This allows a transition region from a condensate with one value of $a(r)$ to a different condensate with another constant value. The Yang-Mills Maxwell equations guarantee that this transition region is necessarily characterized by a topological charge density

$$L(0^{-+}) = -\frac{1}{r^2}[(1-a^2)A_0]' \tag{4.1}$$

To express this Lorentz-invariant density in a convenient form for specific calculations we can use the different formulations

$$E_i^a B_i^a = E_L B_L + 2 E_S B_S = -\frac{1}{r^2}(E_L + (raE_S)') = \frac{1}{r^2} K_1' \tag{4.2}$$

The scalar Lorentz-invariant density in the transition region is given by

$$L(0^{++}) = (A_0')^2 - (\frac{a^2-1}{r^2})^2 + 2\frac{a^2}{r^2}[A_0^2 - (A_1 - \omega')^2]. \tag{4.3}$$

Because of the color gradients, the field-strength densities are enhanced in the domain wall providing a "shrink-wrap" pressure on the interior volume. We can demonstrate the specific properties for these two solitons. The sketch shown in Fig. 3 provides some geometrical scaffolding to the topological structure. For each of these cases we assume that the interior volume with $r \in (0, R_0 - \Delta)$ and consists of a chiral glass condensate.

## The 't Hooft –Polyakov Monopole Soliton

The 't-Hooft-Polyakov soliton has been extensively studied. Specifically, the review by Rossi [22] combines a detailed analysis of the topological structure of non-Abelian monopoles with an expansive study of the fermionic degrees of freedom that can occur. For the 't Hooft-Polyakov soliton, the transition volume with $r \in (R_0 - \Delta, R_0 + \Delta)$ is defined by

$$a(r) = a_{tP}(z) = \frac{1}{2}[1 - \tanh(\kappa z)]$$

$$z = \frac{r - R_0}{\Delta} \quad \frac{\kappa}{\Delta} \gg 1. \tag{4.4}$$

The exterior region given by $r \in (R_0 + \Delta, \infty)$ has the properties

$$a(r) = 0 \quad B_L = -\frac{1}{r^2} \tag{4.5}$$

The stability of the soliton under the requirements (2.5) and (2.7) balancing the external pressure gives the constraint,

$$-\int_{R_0}^{\infty} r^2 dr B_L^2 = -\int_{R_0}^{\infty} \frac{dr}{r^2} = \frac{1}{R_0} \tag{4.6}$$

This constraint specifies the size of the system. The total extended volume describing a single hadron for this spherically-symmetric construction has the topology {T3(color-glass condensate)#[S2xI](CP-odd domain wall)# T3('tHooft Polyakov condensate}. For the solutions describing these solitons, the exterior volume suppresses all color but, importantly, does not necessarily suppress the transmission of a color-neutral Abelian gluon from the interior color-glass condensate,

$$\vec{P}_a(r,t) = \langle A_0 A_1 \rangle \hat{r}_a \exp[i\kappa_0 (r-t)]. \tag{4.7}$$

Therefore, some additional mechanism must be present to guarantee the presence of a mass gap in this situation where internal energy can radiate. In addition, the fact exterior vacuum state contains a radial magnetic field introduces further complications for the renormalization of the quantum theory. Even after averaging over all directions one cannot guarantee the absence of a cosmological constant. This is a general feature of magnetic vacuum states that display the dual-Meissner property. For these solitons, nature requires a quantum field theory renormalization procedure that both suppresses the transmission of Abelian gluons to allow for a "mass gap" and supplies some additional symmetry to cancel the cosmological constant. The search for this renormalization procedure has proved to be a challenge. Monopole world-lines are essential to approaching these issues in the application of these solitons to the problem of color confinement. For example, the gauge fixing required for the quantum theory suggest that both the Coulomb gauge and the Landau gauge emphasize the Fadeev-Popov operator and depend on the infrared properties of the ghost propagator. At the classical level, however, for the 't Hooft- Polyakov solitons the CP-odd domain wall succeeds in that it is able to effectively confine color charge. However, this soliton does not necessarily represent the color group dependencies of confining dynamics such as N-ality and Casimir scaling that are observed empirically. A lot of theoretical effort has been directed at efforts to understand color confinement consistent with these solitons but none has yet met all of the physics constraints.

## The Chiral Transition Soliton

The chiral transition soliton is defined by the extended system {T3(color glass condensate)#[S2xI](CP-odd domain wall)#T3(chiral- vacuum condensate with opposite chirality). As in the previous system, the interior volume is specified to be a color-glass condensate with

$$a(r) = +1 \quad r \in (0, R_0 - \Delta). \tag{4.8}$$

However, the domain wall in the transition volume for these solitons is defined by the constraints,

$$a(r) = a_{CT}(z) = -\tanh(\kappa z) \quad z \in (-1, 1)$$
$$z = \frac{r - R_0}{\Delta} \qquad \frac{\kappa}{\Delta} \gg 1. \tag{4.9}$$

Because $a(r)$ has transformed from $a(r) = +1$ to $a(r) = -1$ through the transition region, the exterior volume is now a chiral vacuum condensate with chirality opposite the chirality of the interior color glass condensate. Note that these solitons are connected to the 't Hooft-Polyakov solitons by

$$a_{CT}(z) = a_{tP}(z) + \tilde{a}_{tP}(z)$$
$$\tanh(-\kappa z) = \frac{1}{2}[1 + \tanh(-\kappa z)] + \frac{1}{2}[-1 + \tanh(-\kappa z)].$$
(4.10)

Therefore, the chiral transition solitons contain an imbedded dyonic color charge in the transition volume of the domain wall. This charge occurs at the surface where the function $a(r)$ passes through $a(r) = 0$. As is the case with the 't-Hooft-Polyakov solitons, the CP-odd layer of topological charge that appears in the transition region of the chiral transition solitons is an essential feature of solutions to the chromostatic Yang-Mills Maxwell equations. A sketch showing typical Lorentz densities appearing in chiral transition solitons is shown in Fig. 5. It should be noted that the solutions leading to chiral transition solitons differ in significant ways from solutions leading to 't Hooft- Polyakov solitons. A chiral transition soliton changes the chirality of the gluonic degrees of freedom so that a change of the radial component angular momentum of $\pm 2$ must occur for a gluon passing through the transition region. The confinement of color for these solitons is connected to this suppression of gluonic transmission.

As discussed in [17] the boundary conditions for the CP-odd structure in the transition region of a chiral transition soliton requires a change of the gauge angle, $\omega(r)$ by

$$\partial \omega(r) = \pm \pi$$
(4.11)

during an excursion through this region from $r \leq R_0 - \Delta$ to $r \geq R_0 + \Delta$. This shows that chirality for massless gluons can generate an extended system requiring a separate topological structure. Both these solitons can be identified as solitons existing in a 1+1 dimensional space-time with a curved metric for a gauge theory coupled to scalar fields, $\Phi(r,t)$. The transverse degrees of freedom play the role of complex scalar fields by making the identification,

$$\Phi(r,t) \exp(i\omega(r,t)) = \Phi_{ia}(r,t)[\varepsilon_{ia}^S(\omega(r,t)) + i\varepsilon_{ia}^A(\omega(r,t))]$$
(4.12)

with

$$|\Phi(r,t)|^2 = \frac{a^2(r,t)}{r^2}$$
(4.13)

This identification is crucial for understanding the implications of the Strong Conjecture. As long as the interior volume of the solitons transforms as an irreducible representation of the Lorentz group we can identify the structures of a 3+1 dimensional field theory with a 1+1 dimensional field theory with spherical symmetry. We now turn to a discussion of the Strong Conjecture Prize.

Section 5.  The Strong Conjecture Prize

The arguments presented in Sec. 4 explicitly illustrate how the color gradients in the SU(2) Yang-Mills Maxwell equations can provide the domain wall of CP-odd topological charge specified in the strong conjecture.

***The confinement mechanism for QCD involves a domain wall of topological (CP-odd) charge separating the interior volume of hadrons from an exterior volume.***

We have, in this article, explored two explicit examples of solitonic solutions to color confinement, the 't Hooft-Polyakov soliton and the chiral transition soliton, each displaying a domain wall of topological charge with spherical symmetry. We hope that this short presentation corrects a major misunderstanding involving the frequently-made assumption that domain walls are not compatible with Lorentz invariance.  This is, of course, not true for spherically symmetric domain walls of finite size with appropriate attention to their interior volume and pressure.  Such extended structures are compatible with Lorentz invariance whenever the interior volume is in an irreducible representation of the rotation group.  Just as we can have commuting and anti-commuting algebras that solve theYang-Mills Maxwell equations we can have solitonic (confining) solutions to the nonlinear equations reflecting these algebras as well as non-solitonic solutions.   Solutions containing a CP-odd domain wall lead directly to color confinement.   These observations illustrate the serious misunderstanding about confinement of non-Abelian color charge that is expressed explicitly in Michael Douglas's [23] report on the status of the Yang-Mills Millenium Prize Problem. In that, otherwise excellent, introduction to the problem he states **"Among the qualitative properties of these solutions, one with some analogies to the 'mass gap' problem would be to establish or falsify the existence of solitonic solutions, whose energy density remains localized for all times (in fact these are not believed to exist)"**.  We have shown explicitly that his statement ignores the nonlinear nature of the non-Abelian systems with color sources and is, thus, only true for self-dual solutions.  It therefore also ignores the very significant role of a gauge connection in providing the nonlinear chiral scaffolding that distinguishes between electric and magnetic degrees of freedom and allows for the emergence of "electric color charge" confinement in classical theory to be connected to rotational invariance and the quantization of angular momentum. Quantum effects are important because in the classical limit, confinement merely involves flux confinement and the study of the Yang-Mills Maxwell equations presented here ignores important quantum effects such as renormalization and the inclusion of both bosonic and fermionic degrees of freedom.  These will be necessary for the proof of the Jaffe-Witten Millenial Prize Problem [24]. However, this does not mean that emergent structures such as domain walls are absent in a quantum theory [25].  In fact, because the reduction of  pure gauge theory in 3+1 dimensions to an Abelian theory in 1+1 dimensions with an r-dependent metric is induced by the imposition of spherical symmetry, the work of Brydges, Frohlich and Seiler [26]on the quantization of the 2-dimensional Abelian Higgs model is completely relevant for the confinement of pure non-Abelian gauge theories with spherical symmetry. The existence of a mass gap in that theory also leads to the existence of a mass gap in the gauge theory. The many problems of taking classical solutions to the Yang-Mills Maxwell equations and turning them into extended structures of a quantum system are not trival. Starting from the premise "Quantum fields are not completely understood either as laws of nature or from a mathematical point of view" Arthur Jaffe [27] has defined a <u>constructive field theory program</u> whose main goals are to:

1. quantize classical field equations,
2. prove the existence of solutions to equations in terms of quantum fields,
3. establish the properties of these solutions.

The strong conjecture allows for a phenomenological approach to the study of color confinement that needs to be augmented by an understanding of quantum effects as described in Jaffe' program.  We believe the study of constructive field theory should include the study of extended topological solitons. As discussed above, to establish the existence of an S-Matrix for extended systems such as hadrons containing confined color sources, we have only to require that the interior volume of each hadron represent an irreducible representation of the rotation group so that the intrinsic angular momentum of each hadron can be quantized.  For this reason, the author has funded a prize of $50,000.00 payable to anyone who can disprove the strong conjecture.  The proof need not be formulated in terms of the axioms established for quantum field theory.  This is a "Physics Prize" not a "Mathematical Prize".  There are many experimental results or theoretical calculations that can, in principle, show the presence or the absence of a domain wall of topological charge.  There are also many differences for the two types of domain wall discussed here.  This is just an introduction to the underlying concepts and a prelude to further work on the subject of color confinement.  We have seen that a soliton containing a concave color mirror combined with concave spatial mirror can confine color if the vacuum outside is "just right".  However an alternative "double-flip" soliton can confine color from a sterile chiral vacuum.

**Acknowledgement**

The author would like to acknowledge the intrepid explorations of his thesis advisor, Geoffrey Chew, in the pursuit of bootstraps and of the ultimate description of the hadronic S-Matrix.

FIGURE CAPTIONS

Fig. 1 This sketch shows a cut-away view of three spherically symmetric volumes in a Euclidean-metric 3-space.
   1. An interior volume (shown in red)
   2. A transition volume (shown in yellow)
   3. An exterior volume (shown speckled).

The total volume has the topological structure T3 (interior volume)# [S2xI] (transition volume)#T3 (exterior volume). To describe the properties required for color confinement we assume the interior volume contains color charges in an adjoint representation of SU(2) and that the exterior volume is empty of color charge. We then show that the field-strength equations for an SU(2) field theory, the Yang-Mills Maxwell equations, require that the transition volume contains a domain wall with CP-odd "topological" charge $E_i^a B_i^a \neq 0$.

Fig. 2 This illustration shows how spherical symmetry can be imposed on an SU(2) gauge theory by parameterizing the gauge connection defined in (3.1) so that the spherical basis vectors $(\hat{r}_i, \hat{\theta}_i, \hat{\phi}_i)$ in Euclidean 3-space are aligned with the adjoint vectors $(\hat{r}_a, \hat{\theta}_a, \hat{\phi}_a)$ in SU(2) color space. The product space then decomposes into longitudinal and transverse components, $\rho_{ia} = \hat{r}_i \hat{r}_a \quad \delta_{ia}^T = (\hat{\theta}_i \hat{\theta}_a + \hat{\phi}_i \hat{\phi}_a) \quad \varepsilon_{ia}^T = (\hat{\phi}_i \hat{\theta}_a - \hat{\theta}_i \hat{\phi}_a)$. The transverse components are distinguished by being either symmetric, ($\delta_{ia}^T$), or antisymmetric, ($\varepsilon_{ia}^T$), on the interchange of color for space indices. In this basis, a rotation in Euclidean 3-space can be compensated by a gauge transformation in adjoint color space. Note that the transverse components of the field-strength tensor given in (3.2) are normalized by the function $a(r,t)$ and include a spatial factor $(1/r)$ compared to the longitudinal Electric field, $E_L(r,t)$, while the longitudinal magnetic field $B_L(r,t)$ is normalized by the factor $\frac{a^2(r,t)-1}{r^2}$.

Fig.3 The transition volume extends from $r = R_0 - \Delta$ to $r = R_0 + \Delta$.

Fig.4 Charge densities for a sample solution involving the 't-Hooft-Polyakov soliton.

Fig.5 Charge densities for a sample solution involving the Chiral Transition soliton.

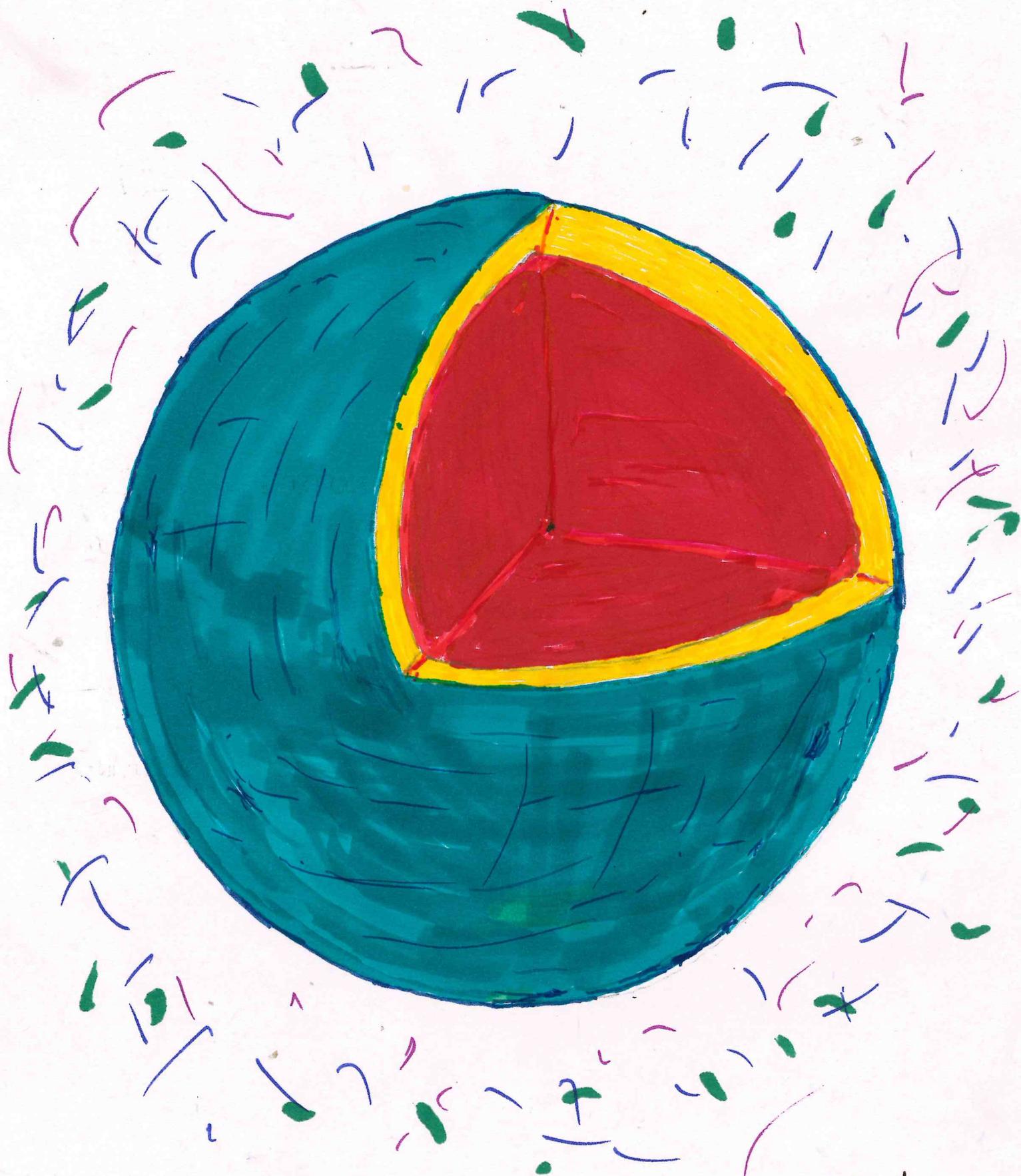

FIG. 1

FIG. 2

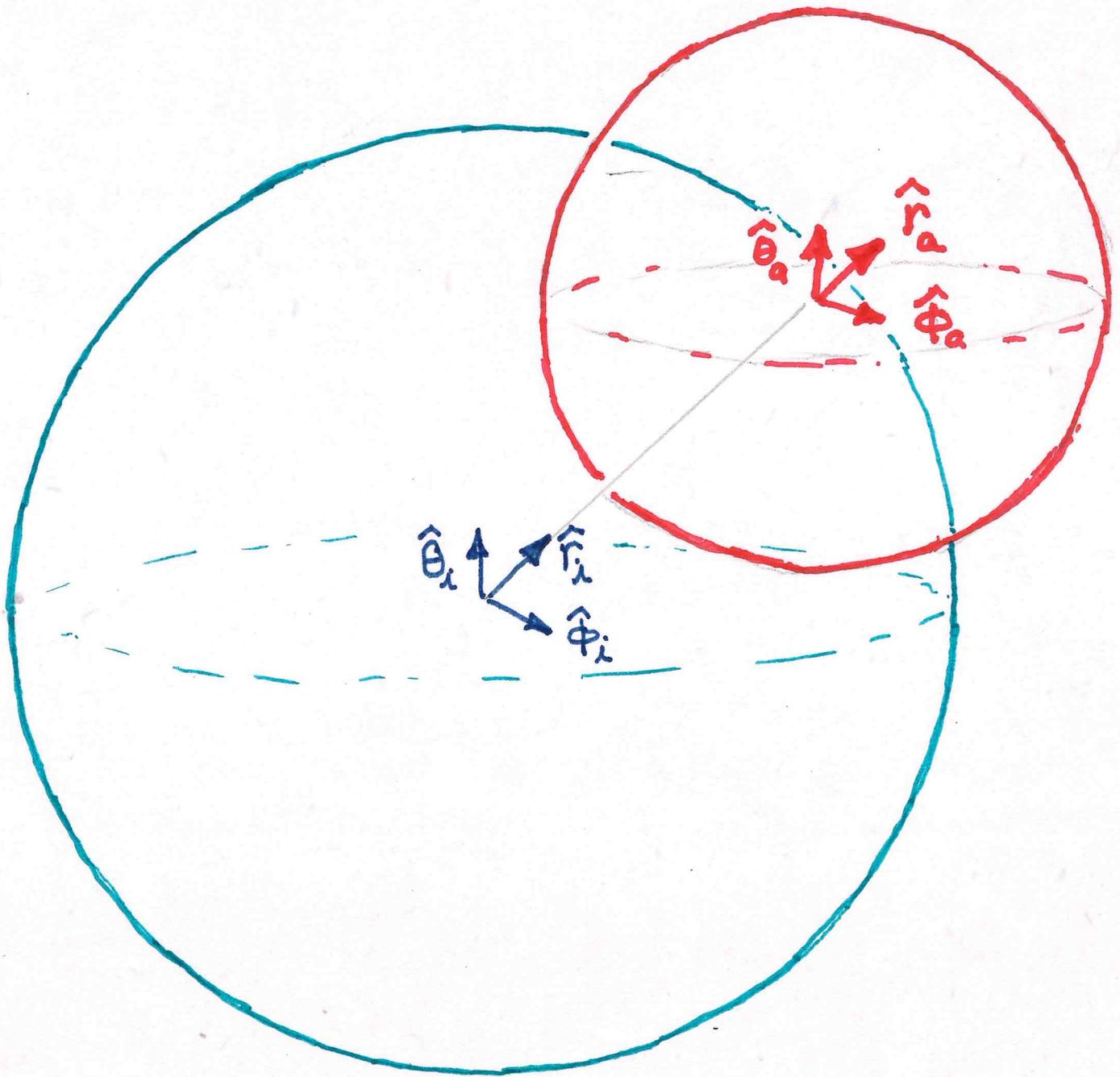

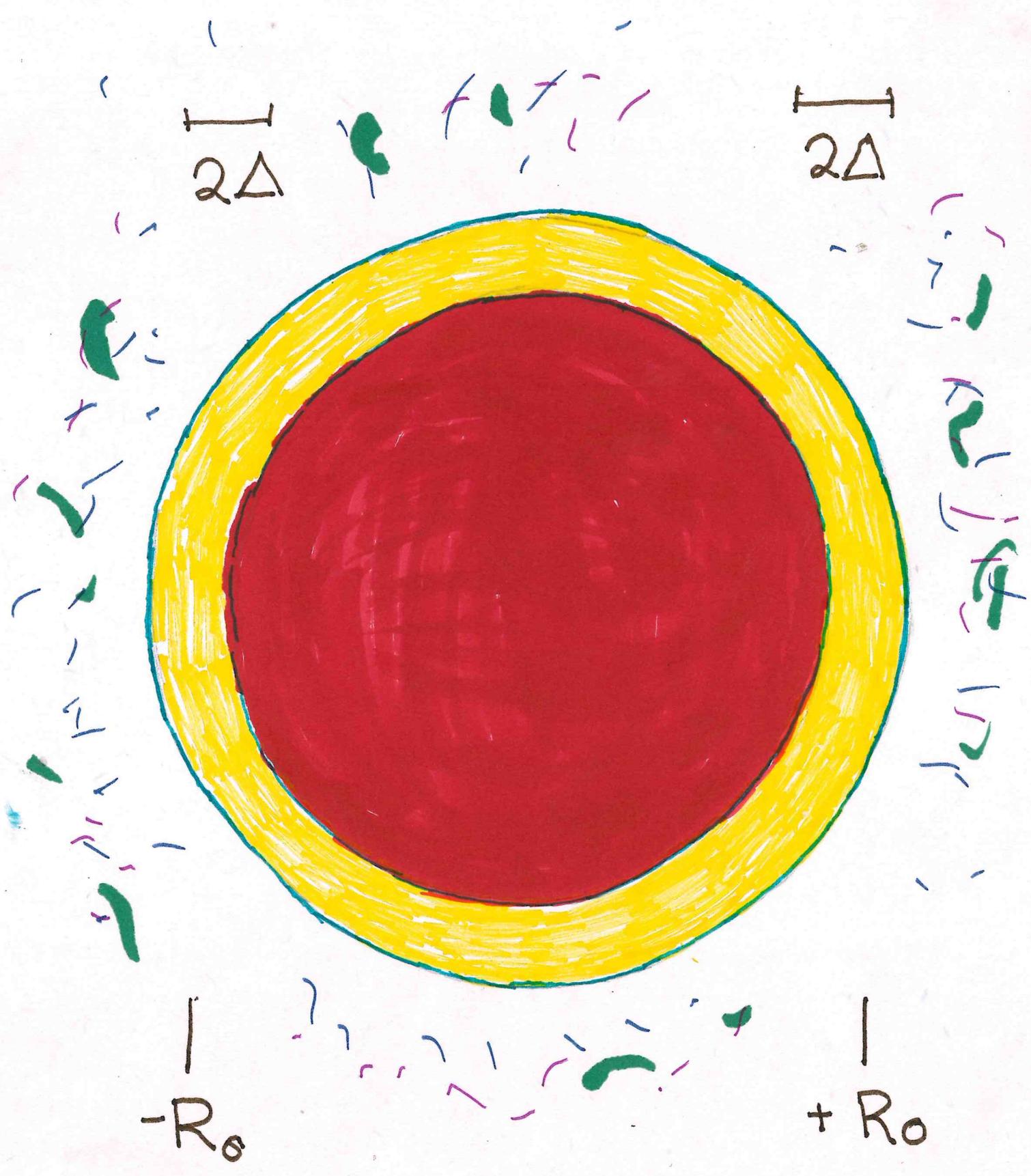

FIG. 3

# 't-Hooft Polyakov Soliton

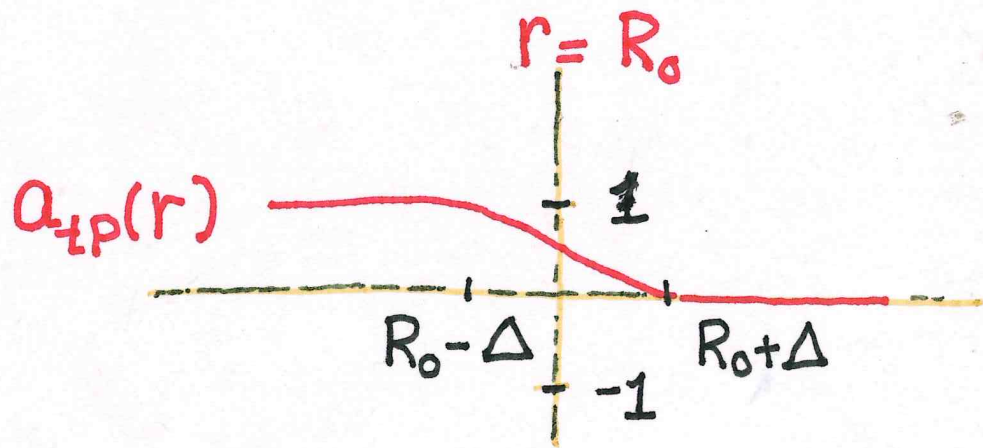

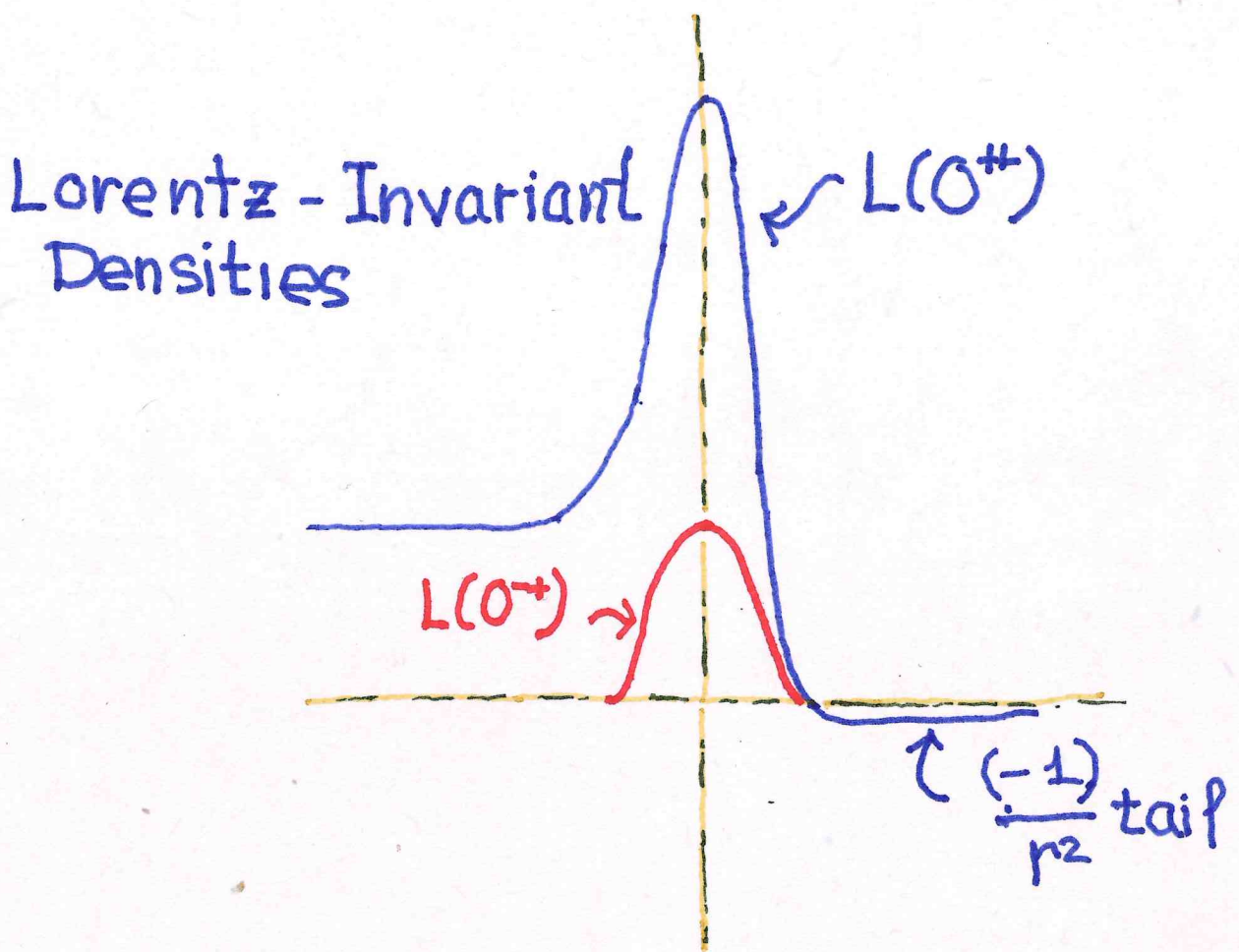

Lorentz-Invariant Densities

$L(0^{++})$

$L(0^{-+})$

$\frac{(-1)}{r^2}$ tail

FIG. 4

# Chiral-transition Soliton

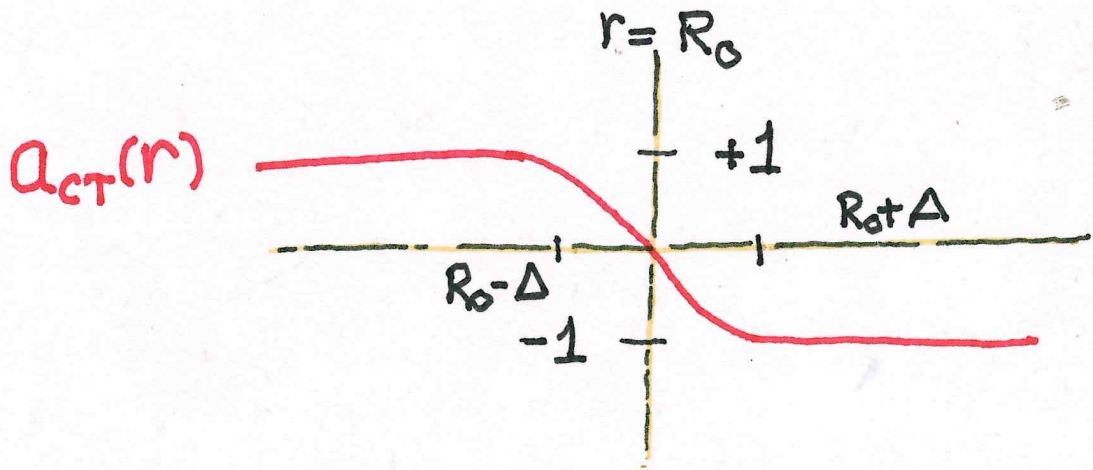

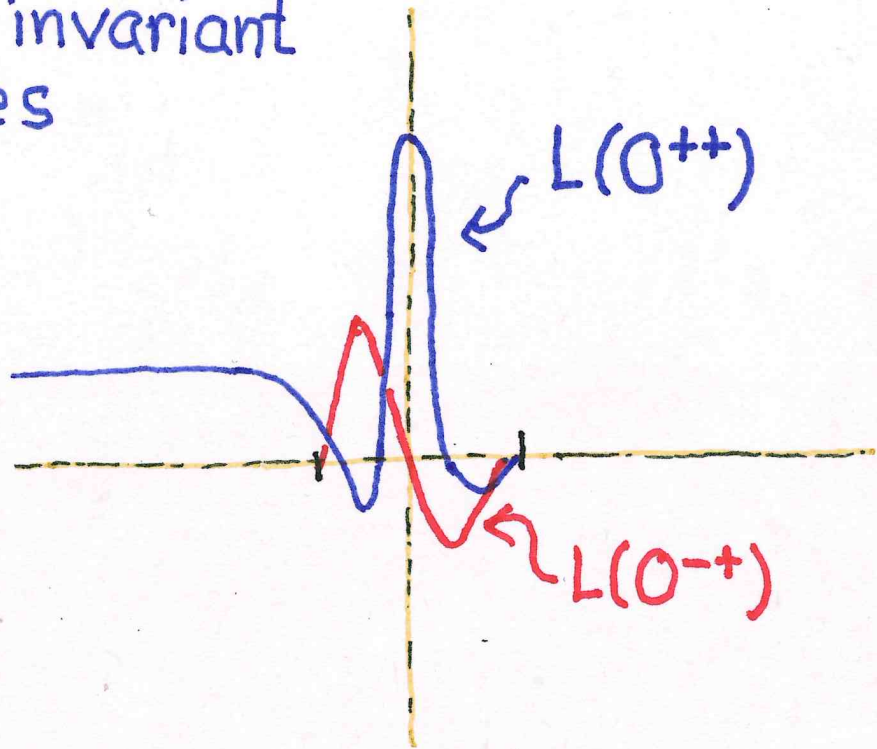

Lorentz-invariant densities

FIG. 5